\documentclass[12pt,preprint]{aastex}

\def\beq{\begin{eqnarray}}
\def\eeq{\end{eqnarray}}

\begin{document}

\title{Vertical Structure of Neutrino-Dominated Accretion Disk
and Applications to Gamma-Ray Bursts}

\author{Tong Liu\altaffilmark{1}, Wei-Min Gu\altaffilmark{2},
Zi-Gao Dai\altaffilmark{1}, and Ju-Fu Lu\altaffilmark{2}}

\altaffiltext{1}{Department of Astronomy, Nanjing University,
Nanjing, Jiangsu 210093, China; tongliu@nju.edu.cn}

\altaffiltext{2}{Department of Physics and Institute of Theoretical
Physics and Astrophysics, Xiamen University, Xiamen, Fujian 361005,
China; guwm@xmu.edu.cn}

\begin{abstract}

We revisit the vertical structure of neutrino-dominated accretion
flows in spherical coordinates. We stress that the flow should be
geometrically thick when advection becomes dominant. In our
calculation, the luminosity of neutrino annihilation is enhanced by one
or two orders of magnitude. The empty funnel along the rotation axis
can naturally explain the neutrino annihilable ejection.

\end{abstract}

\keywords {accretion, accretion disks - black hole physics - gamma
rays: bursts}

\section{Introduction}

Gamma-Ray Bursts (GRBs) are short-lived bursts of gamma-ray photons
occurring at cosmological distances. GRBs are usually sorted of two classes
(Kouveliotou et al. 1993): short-hard GRBs ($T_{90} < 2 \rm s$) and
long-soft GRBs ($T_{90}> 2 \rm s$). The likely progenitors are the
merger of two neutron stars or a neutron star and a black hole
(Eichler et al. 1989; Paczy\'{n}ski 1991; Narayan et al. 1992) and
collapsar (Woosley 1993; Paczy\'{n}ski 1998), respectively. The
popular model of the central engine, namely neutrino dominated
accretion flows (NDAFs), involves a hyperaccreting black hole
with mass accretion rates in the range of
$0.01 \sim 10 M_\odot {\rm s}^{-1}$.
Such a model has been widely investigated in the past decade
(see, e.g., Popham et al. 1999;
Narayan et al. 2001; Kohri \& Mineshige 2002; Di Matteo et al. 2002;
Rosswog et al. 2003; Kohri et al. 2005; Lee et al. 2005; Gu et al.
2006; Chen \& Beloborodov 2007; Liu et al. 2007; Kawanaka \&
Mineshige 2007; Janiuk et al. 2007; Liu et al. 2008). The model can
provide a good understanding of both the energetics of GRBs and the
processes of making the relativistic and baryon-poor fireballs by
neutrino annihilation or magnetohydrodynamic processes (see, e.g.,
Popham et al. [1999] and Di Matteo et al. [2002] for references).

In cylindrical coordinates ($R$, $z$, $\varphi$), Gu \& Lu (2007)
discussed the potential importance of taking the explicit form of
the gravitational potential for calculating slim disk
(Abramowicz et al. 1988) solutions, and
pointed out that the H\={o}shi form of the potential (H\={o}shi
1977), \beq \psi (r,z) \simeq \psi (r,0) + \frac{1}{2} \Omega_{\rm
K}^2 z^2 \ , \eeq is valid only for geometrically thin disks with
$H/R \la 0.2$. Thus the well-known relationship $c_s/\Omega_{\rm K}
H =$ constant does not hold for slim disks with $H/R \la 1$, where
$c_s$ is the sound speed, and $\Omega_{\rm K}$ is the Keplerian
angular velocity. Moreover, with the explicit form of the
gravitational potential, Liu et al.~(2008) found that NDAFs have
both a maximal and a minimal possible mass accretion rate at their
each radius, and presented a unified description of all the three
known classes of optically thick accretion disks around black holes,
namely Shakura-Sunyaev disks (Shakura \& Sunyaev 1973),
slim disks, and NDAFs. These works
are, however, based on the following simple vertical hydrostatic
equilibrium: \beq \frac{1}{\rho} \frac{\partial p}{\partial z} +
\frac{\partial \psi}{\partial z} = 0 \ , \eeq instead of the general
form (Abramowicz et al. 1997): \beq \frac{1}{\rho} \frac{\partial
p}{\partial z} + \frac{\partial \psi}{\partial z} + v_R
\frac{\partial v_z}{\partial R} + v_z \frac{\partial v_z}{\partial
z} = 0 \ , \eeq where $\rho$ is the mass density, $p$ is the
pressure, $v_R$ is the cylindrical radial velocity, and $v_z$ is the
vertical velocity. Since $v_z$ is not negligible for geometrically
thick or slim disks, the solutions in Gu \& Lu (2007) and Liu et al.
(2008) are still not self-consistent. Recently, Gu et al. (2009)
revisited the vertical structure in spherical coordinates and showed
that advection-dominated accretion disks should be geometrically
thick rather than being slim. However, the detailed radiative
cooling was not considered in that work, and therefore no thermal
equilibrium solution was established.

The purpose of this paper is to investigate the vertical structure
of NDAFs with detailed neutrino radiation. In section 2, with the
self-similar assumption in the radial direction, we numerically
solve the differential equations of NDAFs in the vertical direction.
In section 3, we present the vertical distribution of physical
quantities and show the geometrical thickness and the energy
advection of the disk. In section 4, we estimate the luminosity of
neutrino annihilation and discuss some applications to GRBs.
Conclusions are made in section 5.

\section{Equation}

We consider a steady state axisymmetric accretion flow in spherical
coordinates ($r$, $\theta$, $\phi$), i.e., $\partial/\partial t =
\partial/\partial \phi = 0$. We adopt the Newtonian potential $\psi = - GM/r$
since it is convenient for self-similar assumption, where $M$ is
the mass of the central black hole. The basic equations of
continuity and momentum are the following (see, e.g., Xue \& Wang
2005; Gu et al. 2009):

\beq\ \frac{1}{r^2} \frac{\partial}{\partial r}(r^2 \rho v_r) +
\frac{1}{r^2 {\rm sin}\theta} \frac{\partial}{\partial \theta} ({\rm
sin} \theta \rho v_\theta) = 0, \eeq\ \beq\ v_r \frac{\partial
v_r}{\partial r} + \frac{v_\theta}{r} (\frac{\partial v_r}{\partial
\theta} - v_\theta) - \frac{{v_\phi }^2}{r} = -\frac{GM}{r^2} -
\frac{1}{\rho} \frac{\partial p}{\partial r}, \eeq\ \beq\ v_r
\frac{\partial v_\theta}{\partial r} + \frac{v_\theta}{r}
(\frac{\partial v_\theta}{\partial \theta} + v_r) - \frac{{v_\phi
}^2}{r} \cot \theta = -\frac{1}{\rho r} \frac{\partial p}{\partial
\theta}, \eeq\ \beq\ v_r \frac{\partial v_\phi}{\partial r} +
\frac{v_\theta}{r} \frac{\partial v_\phi}{\partial \theta} +
\frac{v_\phi}{r} (v_r + v_\theta \cot \theta) = \frac{1}{\rho r^3}
\frac{\partial}{\partial r} (r^3 T_{r \phi}), \eeq where $v_r$,
$v_{\theta}$, and $v_{\phi}$ are the three components of the
velocity. Here, we only consider the $r\phi$-component of the
viscous stress tensor, $T_{r\phi} = \rho \nu r \partial (v_{\phi}/r)
/\partial r$. The kinematic coefficient of viscosity takes the form:
$\nu = \alpha c_s^2  / \Omega_{\rm K}$ (e.g., Narayan \& Yi 1995),
where the sound speed $c_s$ is defined as $c_s^2 = p/\rho$, the
Keplerian angular velocity is $\Omega_{\rm K} = (GM/r^3)^{1/2}$, and
$\alpha$ is a constant viscosity parameter.

To avoid directly solve the above partial differential equations,
some radial simplification is required since our
main interest is the vertical distribution.
Based on the radial self-similar assumption, Begelman \& Meier (1982)
studied the vertical structure of geometrically thick, optically thick,
supercritical accretion disks. Under the same self-similar assumption,
Narayan \& Yi (1995) investigated the vertical structure of optically
thin advection-dominated accretion flows (ADAFs).
In fact, since the well-known self-similar solutions of ADAFs
(Narayan \& Yi 1994), such type of solutions has been widely investigated
for different classes of accretion, such as slim disks (Wang \& Zhou 1999),
convection-dominated accretion flows (Narayan et al. 2000),
NDAFs (Narayan et al. 2001), and accretion flows with ordered magentic
field and outflows (Bu et al. 2009).
Even though the detailed radiation was considered in some works
(e.g., Di Matteo et al. 2002; Chen \& Beloborodov 2007),
and therefore the solutions cannot be regarded as self-similar solutions,
the self-similar assumption was still adopted such that
the original differential energy equation
can be simplified as an algebraic one.
Furthermore, for optically thick flows, Ohsuga et al. (2005)
showed that their simulations are close to the self-similar solutions
of the slim disk model (e.g., the density profile in their Fig.~11 ).
In our opinion, the radial simplification is necessary for the study of
vertical structure and it is a good choice to take the well-known
self-similar assumption.

Similar to Narayan \& Yi (1995), we adopt the following radial self-similar
assumption: \beq\ \rho(r, \theta) \propto
r^{-3/2}, \eeq\ \beq\ c_s (r, \theta),v_r (r, \theta), v_\phi (r,
\theta)\propto r^{-1/2},\eeq\ \beq\ v_\theta (r, \theta)= 0. \eeq
With the above assumption, equations (5-7) can be simplified as
follows: \beq\ \frac{1}{2} {v_r}^2 + \frac{5}{2} {c_s}^2 +
{v_\phi}^2 - r^2 {\Omega_{\rm K}}^2 = 0, \eeq\ \beq\ \frac{1}{\rho}
\frac{d p}{d \theta} = {v_\phi}^2 \cot \theta, \eeq\ \beq\ v_r =
-\frac{3}{2} \frac{\alpha {c_s}^2}{r \Omega_{\rm K}}. \eeq
Integrating equation (4) over angle we obtain the mass accretion
rate, \beq\ \dot{M} = -4 \pi r^2 \int_{\theta_0}^{\frac{\pi}{2}}
\rho v_r \sin \theta d \theta, \eeq where $\theta_0$ is the polar
angle of the surface.

The equation of state is \beq\ p = p_{\rm gas} + p_{\rm rad} +
p_{\rm e} + p_\nu, \eeq where $p_{\rm gas}$, $p_{\rm rad}$, $p_{\rm
e}$, and $p_\nu$ are the gas pressure from nucleons, the radiation
pressure of photons, the degeneracy pressure of electrons, and the
radiation pressure of neutrinos, respectively. Detailed expressions
of the pressure components were given in Liu et al. (2007). We
assume a polytropic relation in the vertical direction, $p = K \rho
^{4/3}$ , where $K$ is a constant.

The energy equation is written as \beq\ Q_{\rm vis} = Q_{\rm adv} +
Q_\nu, \eeq where $Q_{\rm vis}$, $Q_{\rm adv}$, and $Q_\nu$ are the
viscous heating rate per unit area, the advective cooling rate per
unit area, and the cooling rate per unit area due to the neutrino
radiation, respectively. Here we ignore the cooling of
photodisintegration of $\alpha$-particles and other heavier nuclei.
The viscous heating rate per unit volume $q_{\rm vis} = \nu \rho r^2
[\partial (v_{\phi}/r)/\partial r]^2$ and the advective cooling rate
per unit volume $q_{\rm adv} = \rho v_r (\partial e/\partial r -
p/\rho^2 \partial \rho/\partial r)$ ($e$ is the internal energy per
unit volume) are expressed in the self-similar formalism as \beq\
q_{\rm vis} = \frac{9}{4} \frac{\alpha p v_{\phi}^2}{r^2 \Omega_{\rm
K}}, \eeq \beq\ q_{\rm adv} = - \frac{3}{2} \frac{(p-p_{\rm e})
v_r}{r}. \eeq where the entropy of degenerate particles is
negligible. Thus the vertical integration of $Q_{\rm vis}$ and
$Q_{\rm adv}$ are the following: \beq\ Q_{\rm vis} = 2
\int_{\theta_0}^{\frac{\pi}{2}} q_{\rm vis}  r \sin {\theta} d\theta
\ , \eeq \beq\ Q_{\rm adv} = 2 \int_{\theta_0}^{\frac{\pi}{2}}
q_{\rm adv}  r \sin {\theta} d\theta . \eeq The cooling due to the
neutrino radiation $Q_\nu$ can be written as \beq\ Q_\nu = 2
\int_{\theta_0}^{\frac{\pi}{2}} q_\nu  r \sin {\theta} d\theta \ ,
\eeq where $q_\nu$ is the sum of Urca processes, electron-positron
pair annihilation, nucleon-nucleon bremsstrahlung, and Plasmon decay
(see, e.g., Liu et al. 2007). We therefore can obtain
the luminosity of neutrino radiation $L_{\nu}$ by integrating $Q_\nu$.

In our system, we have six physical quantities varying with
$\theta$, i.e., $v_r$, $v_{\phi}$, $c_s$, $\rho$, $p$, and $T$. The
six equations for solving these quantities are equations (11-13),
(15), the polytropic relation, and the definition of $c_s$ ($c_s^2 =
p/\rho$). Three boundary conditions are required to solve the system
since there is one differential equation, and the boundary
$\theta_0$ and the constant parameter $K$ in the polytropic relation
are unknown. Now we have already two boundary conditions, i.e.,
equations (14) and (16), thus one more boundary condition is
required for solving the system, which is set to be $c_s = 0$
(accordingly $\rho = 0$ and $p = 0$, e.g., Kato et al. 2008, p.~244)
at the surface of the disk, i.e., $\theta =\theta_0$. The numerical
method is as follows. For given $\alpha$, $M$, $\dot{M}$, $r$, and a
test $\theta_0$, from the above six equations and two boundary
conditions (except the energy equation, Eq.~[16]), we can
numerically obtain the vertical distribution of the above six
quantities. With equations (19-21), we then check whether equation
(16) is satisfied for the test $\theta_0$. By varying $\theta_0$ we
can find the exact value of $\theta_0$ for which equation (16) is
matched, and therefore we obtain the exact vertical distribution of
all the variables. In our calculations we take $\alpha = 0.1$ and $M
= 3M_\odot$.

\section{Numerical Results}

Figure 1 shows the variations of the density $\rho$, temperature
$T$, electron fraction $Y_{\rm e}$, and radial velocity $v_r$ with
the polar angle $\theta$ for $\dot{M} = 1 M_\odot \rm s^{-1}$. Here
$Y_{\rm e}$ is defined as $Y_{\rm e} \equiv n_{\rm p}/(n_{\rm p} +
n_{\rm n})$, where $n_{\rm p}$ and $n_{\rm n}$ are the total number
density of protons and of neutrons, respectively (e.g., Beloborodov
2003; Liu et al. 2007). The solid, dashed, and dotted lines
represent the solutions at $r/r_g = 10, 40$, and $100$,
respectively. The profiles of $\rho$ and $v_r$ are similar to that
of the optically thin advection-dominated accretion flows (Narayan
\& Yi 1995), i.e., $\rho$ and $v_r$ (the absolute value) decrease
from the equatorial plane to the surface. On the contrary, electron
fraction $Y_{\rm e}$ increases from the equatorial plane to the
surface and approaches $0.5$ near the surface, which means that the
matter is non-degenerate. The vertical distribution of $v_r$, as
shown in Fig.~1($d$), indicates a multilayer flow with the matter
close to the equatorial plane being accreted much faster than that
near the surface.

Figure $2(a)$ shows the variation of the half-opening angle of the
disk $(\pi /2 -\theta_0)$ with radius $r/r_g$, where $r_g =2 GM/c^2$
is the Schwarzschild radius. The solid, dashed, and dotted lines
represent the solutions with $\dot{M} / M_\odot \rm s^{-1} = 0.1,
1$, and 10, respectively. It is seen that, in the inner region of
the disk, the half-opening angle increases as increasing accretion
rates. For $\dot M = 10 M_\odot \rm s^{-1}$, the inner disk is
extremely thick with the half-opening angle is $\sim 1.4$ radian,
which implies that there exists a narrow empty funnel $\sim
20^\circ$ along the rotation axis. Figure $2(b)$ shows the variation
of the energy advection factor $f_{\rm adv}$ ($\equiv Q_{\rm adv} /
Q_{\rm vis}$) with $r/r_g$. It is seen that advection becomes
important in the inner disk for $\dot M \ga 1 M_\odot \rm s^{-1}$.
Comparing Figs. $2(a)$ and $2(b)$, we find that the curves of the
half-opening angle and the advection factor are similar, which
indicates that the geometrical thickness is relevant to the
advection. For $f_{\rm adv} = 0.5$, it is seen from Fig.~2 that the
half-opening angle is around $1.3$ radian. We therefore stress that
NDAFs should be significantly thick when advection becomes dominant,
which is in agreement with Narayan \& Yi (1995) since their
solutions imply that the flows are extremely thick with the
half-opening angle approaching $\pi/2$.

\section{Applications to GRBs}

In our calculations, the inner disk will be quite thick for large
mass accretion rates, $\dot M \ga 1 M_\odot \rm s^{-1}$. Thus the
volume above the disk shrinks and the radiated neutrino density
increases. Accordingly, the neutrino annihilation efficiency also
increases. We have obtained the neutrino luminosity
$L_{\nu}$ (before annihilation), thus the luminosity of
neutrino annihilation $L_{\nu \bar\nu}$ can be roughly evaluated by the
assumption: $\eta \propto V_{\rm ann}^{-1}$ (see, e.g., Mochkovitch
et al 1993), where $\eta \equiv L_{\nu \bar\nu}/L_{\nu}$ is the
annihilation efficiency, and $V_{\rm ann}$ is the volume above the
disk. For a given outer boundary $r_{\rm out}$, we calculate $V_{\rm
ann}$ by integrating the region of $\theta < \theta_0$ and $r <
r_{\rm out}$. The variations $L_\nu$ and $L_{\nu \bar\nu}$ with
$\dot{M}$ are shown in figure~3. The solid lines correspond to the
present solutions whereas the dashed lines correspond to those in
Liu et al. (2007). As shown in Fig.~3, for the same $\dot{M}$,
$L_\nu$ is comparable, whereas $L_{\nu \bar\nu}$ in the present
results is significantly larger than that in Liu et al. (2007) by
one or two orders of magnitude. Moreover, we find that for $\dot{M}
= 5 M_\odot \rm s^{-1}$, $L_{\nu \bar\nu}$ is very close to $L_\nu$,
which means that the density of radiated neutrino is so large that
the annihilation efficiency is close to 1. Thus we can expect that
$L_{\nu \bar\nu}$ is roughly equal to $L_\nu$ for $\dot{M} \ga 5
M_\odot \rm s^{-1}$.

Many previous works have calculated the annihilation
luminosity and claimed that the NDAF mode can provide enough energy
for GRBs. However, GRBs are generally believed to be a jet with a
small opening angle $\theta_{\rm jet}$. The problem is that, the
annihilation could not be limited into such a small angle even
though the region well above the inner disk have larger luminosity
than other place. We argue that our model is preferably to explain
the ejection-like radiation, because the disk is adequately thick
and there exists a narrow empty funnel along the rotation axis,
which can naturally explain the neutrino annihilable ejection.

\section{Conclusions}

In this paper we revisit the vertical structure of NDAFs in
spherical coordinates. The major points we wish to stress are as
follows:
\begin{enumerate}
\item
We show the vertical structure of NDAFs and stress that the flow should
be significantly thick when advection becomes dominant.

\item
The luminosity of neutrino annihilation is enhanced by one or two
orders of magnitude.

\item
The narrow empty funnel ($\sim 20^{\circ}$) along the rotation axis
can naturally explain the neutrino annihilable ejection.

\end{enumerate}

\acknowledgments

We thank Katsuaki Asano, H.-Thomas Janka and Yi-Zhong Fan for
beneficial discussion and comments. This work was supported by the
National Basic Research Program (973 Program) of China under Grant
2009CB824800 (JFL and WMG), the National Natural Science Foundation
of China under grants 10778711 (WMG), 10833002 (JFL and WMG),
10873009 (ZGD), and the China Postdoctoral Science Foundation funded
project 20080441038 (TL).

\newpage

\begin{figure*}
\centering
\includegraphics[angle=0,scale=0.8]{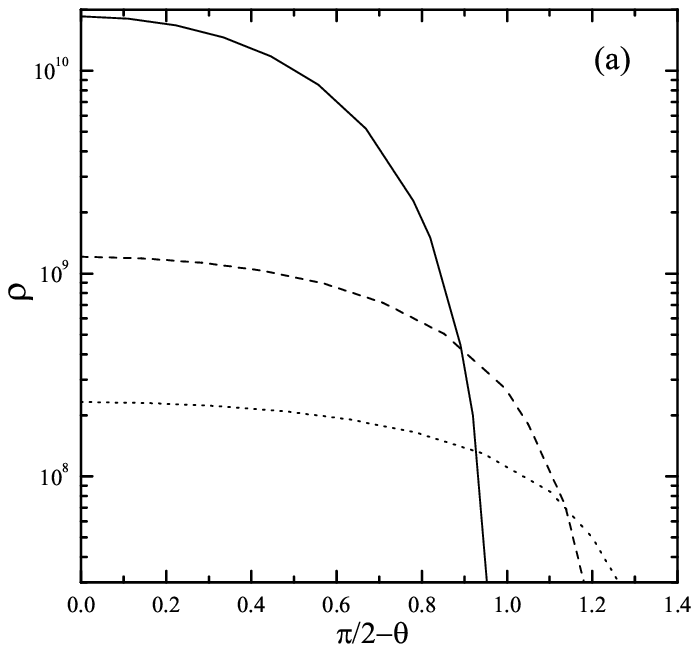}
\includegraphics[angle=0,scale=0.8]{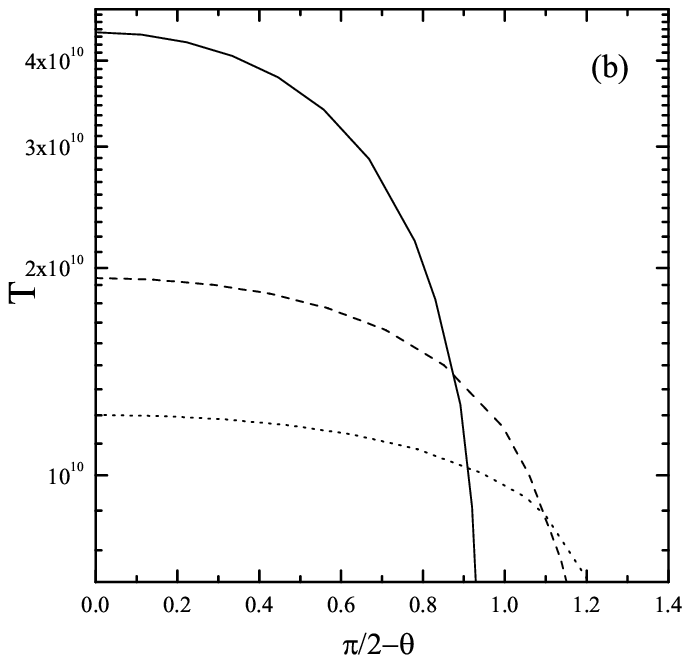}
\includegraphics[angle=0,scale=0.8]{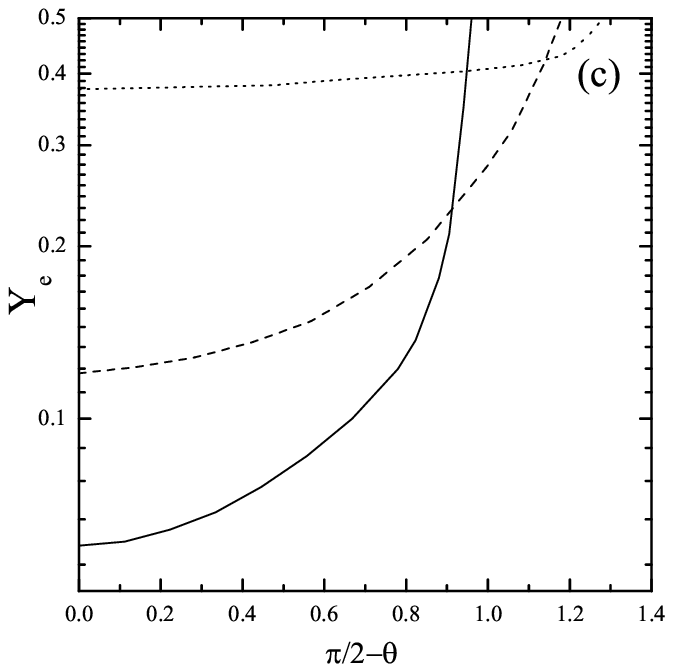}
\includegraphics[angle=0,scale=0.82]{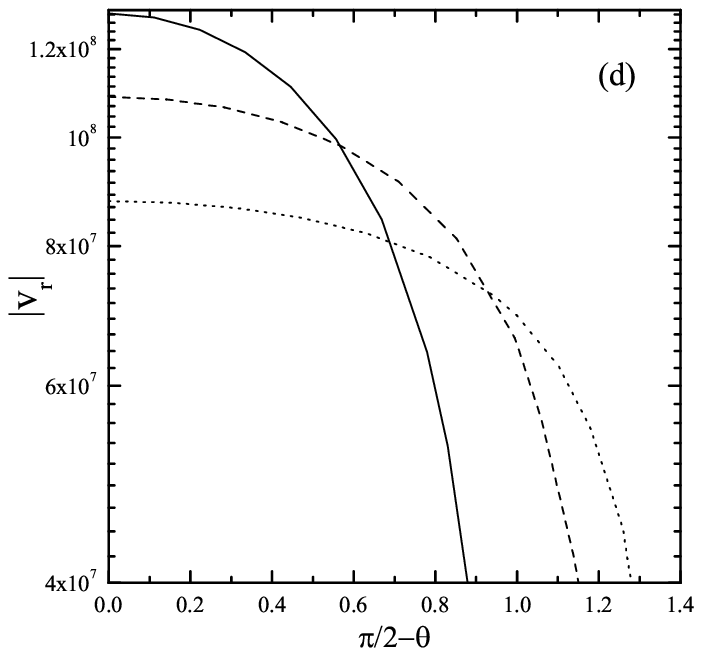}
\caption{ Variations of the density $\rho$, temperature $T$,
electron fraction $Y_{\rm e}$, and radial velocity $v_r$ with the
polar angle $\theta$, for which the given parameters are
$\dot{M}/M_\odot \rm s^{-1} = 1$ and $r/r_g = 10$ (solid lines),
$40$ (dashed lines), $100$ (dotted lines).} \label{sample-figure1}
\end{figure*}

\newpage

\begin{figure*}
\centering
\includegraphics[angle=0,scale=0.8]{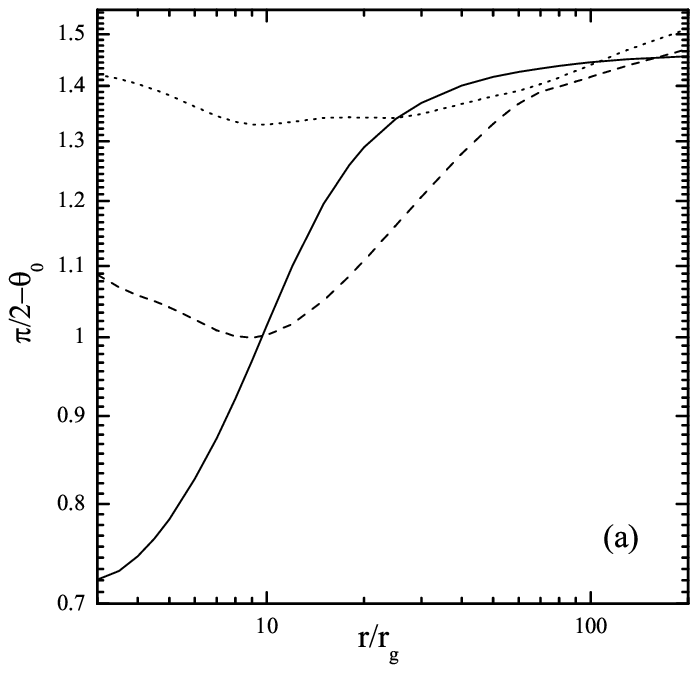}
\includegraphics[angle=0,scale=0.8]{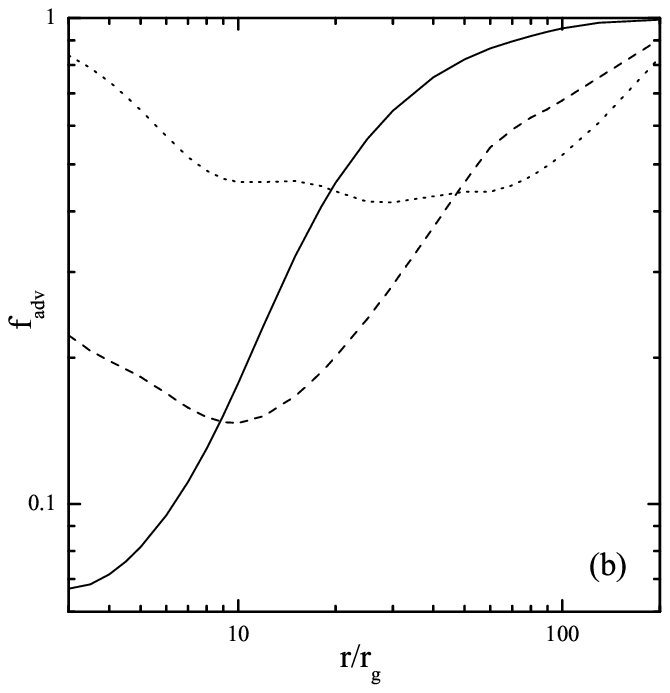}
\caption{Variations of the half-opening angle of the disk $(\pi /2
-\theta_0)$ and the advection factor $f_{\rm adv}$ with radius
$r/r_g$, for which the given parameter is $\dot{M}/M_\odot \rm
s^{-1} = 0.1$ (solid line), $1$ (dashed line), $10$ (dotted line).}
\label{sample-figure2}
\end{figure*}

\newpage

\begin{figure*}
\centering
\includegraphics[angle=0,scale=1.5]{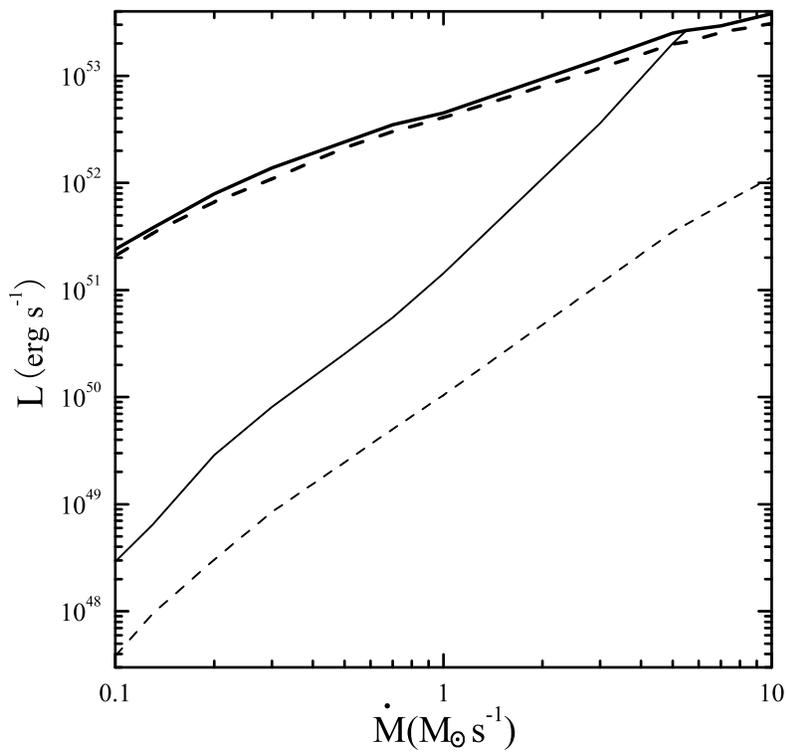}
\caption{Neutrino luminosity $L_\nu$ (thick lines) and
annihilation luminosity $L_{\nu \bar\nu}$ (thin lines) for
varying mass accretion rates $\dot{M}$. The solid lines correspond
to the present solutions, whereas the dashed lines correspond to the
solutions of Liu et al.~(2007).} \label{sample-figure3}
\end{figure*}

\clearpage

\end{document}